\documentclass{mem}
\usepackage{amsmath}
\usepackage{natbib}\usepackage{txfonts}\usepackage{balance}
\usepackage{graphicx}

\idline{0}{1}
\begin{document}
\def\mhyph{\, \mathchar`- \, }
\def\nuc#1{${}^{#1}$}
\def\nucm#1#2{{}^{#1}{\rm #2}}
\def\abra#1#2{[{\rm #1}/ {\rm #2}]}

\newcommand{\feoh}{[{\rm Fe} / {\rm H}]}
\newcommand{\msun}{\, M_\odot}
\newcommand{\Zsun}{\, Z_\odot}
\newcommand{\mmd}{M_{\rm md}}
\newcommand{\dm}{\Delta_{\rm M}}

\title{
Current signatures and search for Pop. III stars in the Local Universe
}

   \subtitle{}

\author{
Y. \,Komiya\inst{1}, 
T. \,Suda\inst{1}\
\and M. \, Fujimoto\inst{2,3}
          }

  \offprints{Y. Komiya}

\institute{
National Astronomical Observatory of Japan, Osawa, Mitaka, Tokyo, Japan
\and
Faculty of Engineering, Hokkai-gakuen University, Sapporo, Hokkaido 062-8605, Japan
\and
Department of Cosmoscience, Hokkaido University, Sapporo, Hokkaido 060-0810, Japan
\email{yutaka.komiya@nao.ac.jp}
}

\authorrunning{Bonifacio }

\titlerunning{Search for Pop. III stars}

\abstract{
Recent numerical studies argue that low-mass stars can be formed even at zero-metallicity environment. 
These low-mass Population III(Pop.~III) stars are thought to be still shining and able to be observed in the Local Universe. 
%We build a new model of chemical evolution based on the hierarchical galaxy formation scenario, and discuss possibility to observe Pop. III survivors and the expected observational features. 
Most low-mass Pop.~III stars are thought to be formed as secondary companions in binary systems. 
They can be escaped from their host mini-halos when their primary companions explode as supernovae. 
In this paper, we estimate the escape probability of the low-mass Pop.~III stars from their host mini-halos. 
We find that $\sim 100$ Pop.~III stars are expected. 
We also compute spatial distribution of these escaped Pop.~III survivors by means of the semi-analytic hierarchical chemical evolution model. 
Typically, they are distributed around $\sim 2$Mpc away from the Milky Way but 5 -- $35\%$ of the escaped stars fall into the Milky Way halo. 
These escaped Pop.~III stars are possibly detected by very large scaled surveys being planned. 
}
\maketitle{}

\section{Introduction}

In spite of the longstanding effort, population III (Pop. III) stars, which are stars without metal, have not been discovered yet. 
Extremely metal-poor (EMP) stars in the Milky Way halo are searched by the large scaled surveys; HK survey, HES survey, and SDSS/SEGUE. 
These surveys found thousands of EMP stars with $\feoh<-3$ but no star below $\feoh = -6$. 
The observed most metal-poor objects, which are refferd to as hyper metal-poor (HMP) stars, have metallicity between $\feoh = -5$ to $-6$ 
(HE1327-2326, $\feoh=-5.7$, \citet{Frebel05}; HE0107-5240, $\feoh=-5.4$, \citet{Christlieb02}). 

%(theory)
On the other hand, from a theoretical perspective, the possibility of the formation of low-mass Pop.~III stars has been pointed out. 
Recent numerical simulations argue that a primordial gas cloud fragments into multiple pieces,  
and form a Pop.~III binary \citep{Machida08, Turk09, Stacy10} or a Pop.~III star cluster \citep{Clark08, Clark11, Greif11}. 
These results indicate that low-mass stars can be formed even in the pristine environment unaffected by prior star formation. 
Typical mass of the second generation Pop.~III (Pop.~III.2) stars, which are born from the gas ironized by the first generation stars, are thought to be smaller than the first generation Pop.~III (Pop.~III.1) stars because of more efficient cooling \citep{Uehara00, OShea05}. 
If Pop.~III stars with mass $m \lesssim 0.8 \msun$ are formed, they are still shining by nuclear burning to data and can be observed in the Local Universe. 
We refer to these low-mass Pop.~III stars which survive to data as Pop.~III survivors. 

%(in MW halo)
In our previous studies, we investigated the change of the surface abundance by accretion of interstellar medium (ISM) on Pop.~III survivors in the Milky Way halo \citep{Komiya09L, Komiya10}. 
We found that the surfaces of Pop.~III survivors are polluted to $\feoh \sim -5$. 
This result indicates that Pop.~III survivors are observed as HMP stars when they are in the Milky Way halo. 
HMP stars are possibly Pop.~III survivors but there are some other scenarios in which HMP stars are formed as chemically second generation stars \citep{Umeda03, Iwamoto05, Maynet06}. 

%(escape scenario)
Is it impossible to observe the unpolluted Pop.~III survivors with zero surface metallicity in the Local Universe?
In this paper, we propose that stars with $Z=0$ can be found {\it outside} the Milky Way halo. 

%(escape scenario)
In the $\Lambda$CDM cosmology, Pop.~III stars should be formed in the mini-halos with total mass $M_{\rm h} \sim 10^6\msun$ \citep[e.g.][]{Yoshida03}. 
Low-mass Pop.~III survivors have possibilities to escape from their host mini-halos
because of the small gravitational potential of mini-halos in which Pop.~III stars are formed. 

Two channels to release Pop.~III survivors can be considered. 
If a low-mass Pop.~III star is formed as a secondary of a binary with a massive primary star, it will go out from the binary system when the primary explode as a SN. 
The other channel is the escape of a low-mass star through gravitational slingshot from a Pop.~III star cluster. 
These stars can go away from mini-halo, and stay in intergalactic space. 
These ``escaped Pop.~III'' stars are free from the surface pollution because it is difficult to accrete the metal enriched gas in the intergalactic space. 

In this paper, we investigate the binary escape scenario. 
We evaluate the escape probability of Pop.~III and EMP survivors from their host mini-halo and estimate the number and spatial distribution of them around the Milky Way at $z=0$. 
We also discuss the detection probability of the escaped Pop.~III survivors. 

\section{Escape of Pop. III stars from their host mini-halo}

%(escape criterion)
When the primary star of a binary explode as a SN, 
the secondary star can be released from the binary system 
since the gravitational potential suddenly decreases. 
If the kinetic energy of the secondary star is larger than the sum of the gravitational potential of its host mini-halo and of the remnant object after the SN explosion, 
\begin{equation}
\frac{1}{2}v_{\rm orb}^2 - \frac{G m_{\rm rem}}{r_{\rm bin}} + \Psi_{\rm halo}(M_{\rm h}, z) > 0, 
\label{criterionEq}
\end{equation}
the secondary star escape from the mini-halo,  
where 
$v_{\rm orb}$ and $r_{\rm bin}$ are the velocity of the secondary star and the separation of the binary, respectively. 
$m_{\rm rem}$ is the mass of the remnant (neutron star or black hole) of the primary star, 
and $\Psi_{\rm halo}$ is the gravitational potential of the mini-halo as a function of mini-halo mass and redshift. 
The terminal velocity of the escaped star becomes 
\begin{equation}
v_{\rm esc} = \left( v_{\rm orb}^2 - \frac{2 G m_{\rm rem}}{r_{\rm bin}} + 2 \Psi_{\rm halo}(M_{\rm h}, z) \right)^{1/2}. 
\end{equation}

\begin{figure}[]
\resizebox{\hsize}{!}{\includegraphics[clip=true]{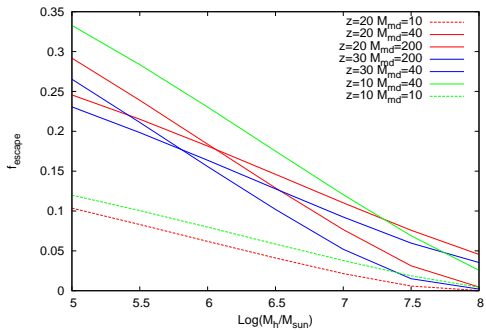}}
\caption{
Escape fraction of low-mass secondary stars of binaries as functions of mass of their host mini-halo. 
From a mini-halo of $m \sim 10^6\msun$, $\sim 20 \%$ of Pop.~III survivors are escaped when the very high-mass IMF is adopted. 
}\label{percent}
\end{figure}

For the gravitational potential of mini-halos, we adopt the NFW profile with the concentration parameter $c=10$. 
It is assumed that Pop.~III stars are formed at the center of the mini-halos. 
$m_{\rm rem}$ is given as a function of initial mass of the primary following \citet{Woosley02}. 
$v_{\rm orb}$ and $r_{\rm bin}$ are computed from mass of primary and secondary, and period of the binary assuming the circular orbit. 

We estimate the escape fraction adopting the following parameter for the distribution of the primary mass, $m_1$, the mass ratio $q=m_2/m_1$, and the binary period. 

%(IMF)
We use the lognormal IMF. 
In our previous studies, we gave constraints on the IMF of EMP stars based on the statistics of carbon-enhanced EMP stars and the total number of observed EMP stars \citep{Komiya07, Komiya09}. 
As a result, we yield the lognormal IMF of medium mass $\mmd = 10 \msun$ and dispersion $\sigma = 0.4$ for EMP stars. 
Theoretical studies argue that the typical mass of Pop.~III stars is larger than the EMP stars. 
We also adopt the very high mass IMF with $\mmd = 40 \msun$ or $\mmd = 200 \msun$ for Pop.~III stars. 

%(binary parameter)
For the mass ratio distribution, we adopt two functions; 
a flat mass ratio distribution, $n(q) = 1$, as a fiducial one
and a bottom heavy distribution, $n(q) = 2(1-q)$.  
For the period distribution of binaries, we use the observational result of \citet{Duquennoy91} as a fiducial one and adopt also one of \citet{Rastegaev10}. 
%We assume circular orbit for all binaries. 
%Remnant mass of primary stars is given as a function of initial mass $m$ by \citet{Woosley}. 

%(result)
Secondaries of close binaries with period smaller than a few hundred yrs escape from mini-halos. 
Figure~\ref{percent} shows the percentage of the escaped stars among Pop.~III survivors as functions of mini-halo mass. 
%In this figure, we plot result of model with $n(q) = 1$
When we assume the very high mass IMF with $\mmd \geq  40\msun$, $\sim 20 \%$ of the low-mass secondaries of Pop.~III binaries escape from mini-halos with $10^6 \msun$. 
In the case of $\mmd = 10\msun$, the percentage becomes less than $10\%$ because majority of primary companion of Pop.~III survivors are intermediate massive stars. 
From more massive mini-halos with $10^8\msun$, the escape frequency is less than $\sim 5\%$ due to their deep gravitational potential. 
%Considerable fraction of Pop.~III survivors escapes from mini-halo.

\section{Distribution of the escaped Pop.~III stars in the Local universe}
In this section, we estimate the total number of the stars escaped from mini-halos as the building-blocks of the Milky Way and the spatial distribution of the escaped Pop.~III or EMP stars. % by the semi-analytic method. 
In our previous paper, we build a merger tree of the Milky Way halo using the extended Press-Schechter method by \citet{SK99}, and follow the star formation history and the chemical evolution along the tree \citep{Komiya10, Komiya11, Komiya13}. 
We use the results of these previous studies. 

%(orbit after escape)
We pick-up the Pop.~III and EMP survivors formed in the hierarchical chemical evolution model, and set the binary parameters randomly following the distribution described above for each star.  
%and compute escape frequency as a function of their formation redshift and the mass of their host mini-halos. 
Stars which satisfy Eq.~\ref{criterionEq} escape from mini-halos. 

We assume that the main halo of the merger tree has been stayed from the beginning at the position where the Milky Way locate now, and other mini-halos were accreted to the main halo, and compute the distance of the escaped Pop.~III stars from the Milky Way. 

For each escaped Pop.~III star, we follow its radial orbit, $r_s(t)$, by integrating the equation of motion, 
\begin{equation}
\frac{d^2 r_{s}}{dt^2} = - \frac{G M_{\rm in}(r_{s},t)}{r_{s}^2} + \frac{\Lambda c^2}{3} r_s + \frac{l_{s}^2}{r_{s}^3} 
\end{equation}
, where $\Lambda$ is the cosmological term 
and $l_{s} \equiv r_{s, 0} v_{\rm ej} \sin\theta$ is the specific angular momentum of the star. 
We set the escape direction, $\theta$, randomly. 
$M_{\rm in}(r_{s}, t)$ is the total mass inside the radius $r_{s}$ at time $t$ and we assume that 
\begin{equation}
M(r_s,t) = \frac{4}{3}\pi r_s^3 \rho_{\rm av}(t) + M_{\rm main}(t)
\end{equation}
, where $M_{\rm main}$ is mass of the main halo and $\rho_{\rm av} $ is the average density of the Universe. 
The initial distance and initial radial velocity are 
\begin{align}
r_{s, 0} &= r_h \\
v_{s, 0} &= v_{\rm ej} \cos \theta + v_h.
\end{align}
, where $r_h$ and $v_h$ are distance and radial velocity of the mini-halo in which the Pop.~III star is formed. 

%(mini-halo spatial distribution)
The distance of the mini-halos is estimated by the following method. 
%We compute the distance from main halo and radial velocity $v_{halo}$ of mini-halos by integrating the above equation. 
In the Press-Schechter formalism, ``halo'' is the collapsed spherical overdense region. 
In the extended Press-Schechter theory, %formation of halos is described as gravitational collapse of spherical density fluctuation. 
``merger'' of two halos is described as a gravitational collapse of the region which contain the both two halos. 
Distance between the two halos which merge to form a halo with total mass, $M_{\rm tot}$, at $t_{\rm m}$ 
is thought to be similar with a radius of the spherical shell of mass $M_{\rm tot}$ which collapse at $t_{\rm m}$. 
Evolution of the spherical shell is simply described by the following equation,  
\begin{equation}
\frac{d^2 r_h}{dt^2} = -\frac{G M_{\rm tot}}{r_h^2} + \frac{\Lambda c^2}{3} r_h . 
\end{equation}
We compute the distance, $r_h$, of a Pop.~III forming mini-halo which merges with the main halo at $t_{\rm m}$ by integrating this equation with conditions of $r_h(t_{\rm m})=0$ and $M_{\rm tot}=M_{\rm main}(t_{\rm m})$.

\begin{figure}
\resizebox{\hsize}{!}{\includegraphics[clip=true]{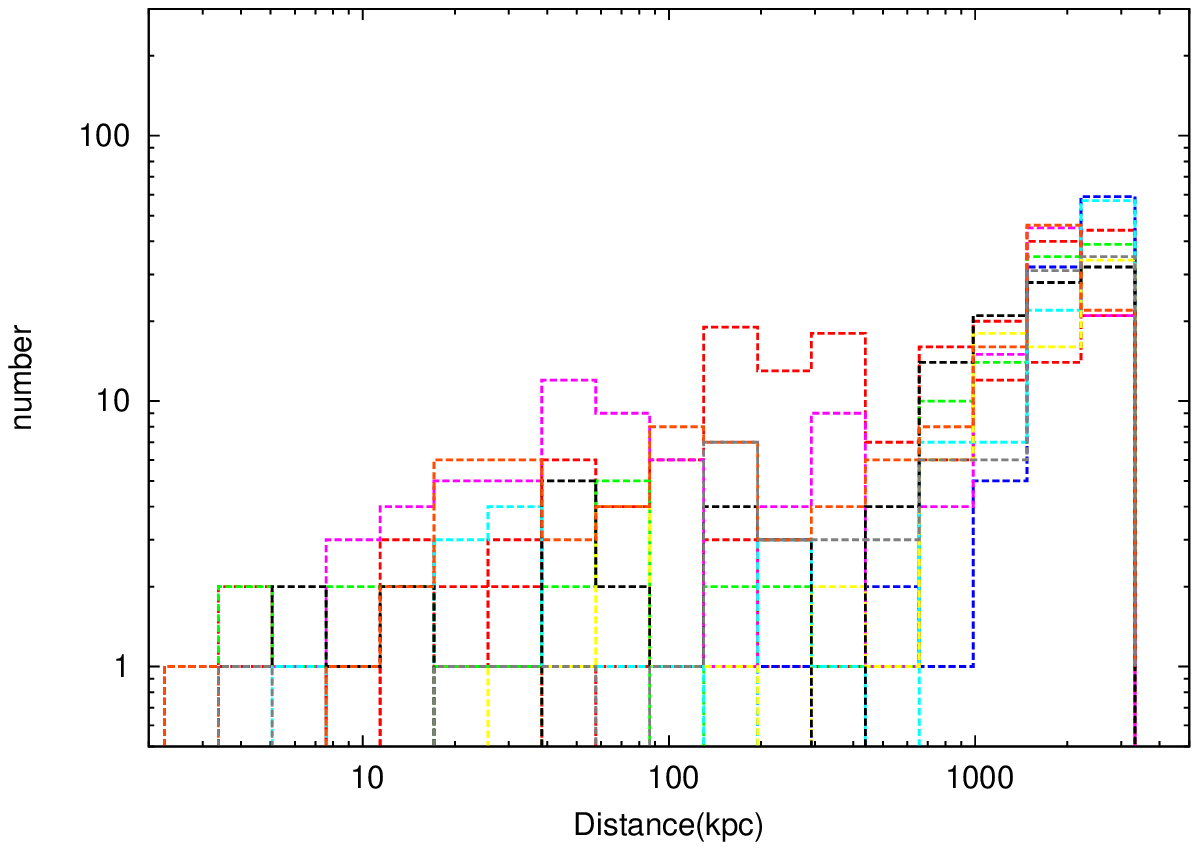}}
\resizebox{\hsize}{!}{\includegraphics[clip=true]{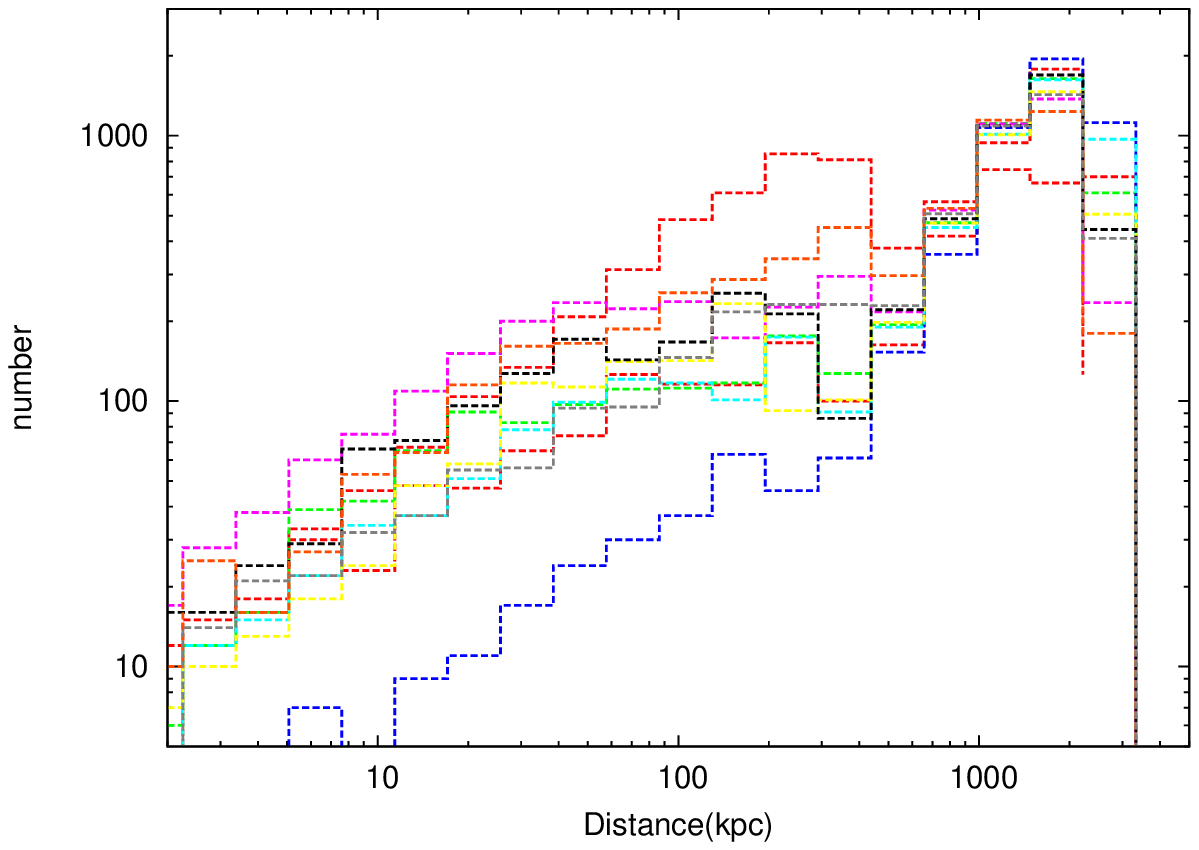}}
\caption{
The predicted distribution of the escaped Pop.~III survivors (top panel) and EMP survivors (bottom panel) around the Milky Way halo. 
%5 -- $35\%$ of the escaped stars fall into the Milky Way halo.
%The other stars are distributed around $\sim 2$Mpc away from the Milky Way. 
We overplot the results of 10 computation runs for different merger trees
}\label{dist}
\end{figure}

\begin{figure}
\resizebox{\hsize}{!}{\includegraphics[clip=true]{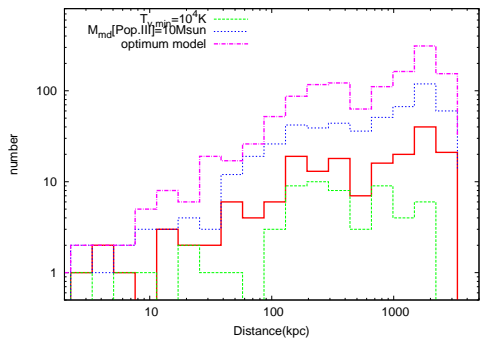}}
\caption{Parameter dependence of the distribution of the escaped Pop.III stars. 
}\label{param}
\end{figure}

%(total number)
In the building blocks of the Milky Way, $\sim 500$ of Pop.~III survivors are formed in total when we adopt $\mmd=40\msun$, $n(q) = 1$ and the period distribution of \citet{Duquennoy91}. 
More than 100 of them are escaped from their host mini-halos. 
%They escape at $z>10$ and $10^6\msun$. 
%(spatial distribution)
They are formed at $z \sim 20$, and escaped after $\sim10^7$yrs with velocity $v_{\rm ej}=5$ -- 70km/s and $v_h=50$ -- 250km/s. 
Top panel of Figure~\ref{dist} shows the predicted distribution of the escaped Pop.~III survivors. 
Typically, they are distributed around $\sim 2$Mpc away from the Milky Way. 
5 -- $35\%$ of the escaped Pop.~III survivors have accreted to the Milky Way and distributed around the virial radius of its dark matter halo. 

%(EMP)
We also compute the distribution of the escaped EMP survivors. 
More than 1000 EMP stars are escaped from mini-halos in our model. 
The distribution of them is similar to the escaped Pop.~III stars, as shown in the bottom panel of Figure~\ref{dist}.

%(parameter dependence)
Figure~\ref{param} shows parameter dependence of the predicted distribution of the escaped Pop.~III stars. 
When we adopt the optimum parameter set; $\mmd=10\msun$, $n(q) = 2(1-q)$, and the period distribution of \citet{Rastegaev10}, the predicted  number of the escaped Pop.~III survivors becomes $\sim7$ times larger than the fiducial case. 
On the other hand, when we assume that stars are formed only in halos with virial temperature higher than $10^4$K, the number of the escaped Pop.~III star reduce to $\sim1/5$. 

For all cases, the escaped Pop.~III stars are distributed around 2Mpc away from the Milky Way. 
This is because the spatial distribution of these stars are dominated by the motion of their host mini-halos rather than the binary properties ($v_h > v_{\rm ej}$) for most the escaped Pop.~III stars. 

%(observation)
The luminosity of Pop.~III survivors at red giant branch or horizontal branch is $\sim100L_{\odot}$ of more. 
The observed magnitude of these stars outside the Milky Way is $\sim26$ -- 28 mag. 
There is a very wide ($\sim1400 {\rm deg}^2$) and deep ($i \lesssim 26$mag) survey using Hyper Suprime-Cam planned. 
Such a large scale survey possibly detects these escaped Pop.~III or escaped EMP survivors.

\section{Summary}
We discuss the observation probability of the Pop.~III stars in the Local Universe. 
We propose that the unpolluted Pop.~III survivors can be observed outside the Milky Way halo. 
When the Pop.~III binaries are formed, $\sim 20\%$ of low-mass secondary stars are thought to be escaped from their host mini-halos when their primaries explode as SNe. 
By means of the semi-analytic model for the star formation history of the Milky Way, we estimate the number and distribution of the escaped Pop.~III survivors. 
We predict that 20 -- 1000 Pop.~III survivors are distributed around $\sim 2$Mpc away from the Milky Way. 
5 -- $35\%$ of the escaped Pop.~III stars fall to the Milky Way and is distributed around the virial radius of the dark halo. 
Thousands of EMP survivors may also be distributed at similar distance. 
The observational magnitude of these stars at $\sim 2$Mpc is 26 -- 28 mag when they are giant stars. 
There is a possibility to observe these stars by very wide and deep survey using such as Hyper Suprime-Cam.

\begin{acknowledgements}
This work was supported by the JSPS KAKENHI Grant Number 25800115, 23224004.
\end{acknowledgements}

\bibliographystyle{aa}

\begin{thebibliography}{}

\bibitem[Christlieb et al.(2002)]{Christlieb02}	Christlieb, N., Bessell, M. S., Beers, T. C., Gustafsson, B., Korn, A., Barklem, P. S., Karlsson, T.; Mizuno-Wiedner, \& M., Rossi, S.\ 2002, Nature, 419, 904
\bibitem[Clark et al.(2008)]{Clark08} 	Clark, P. C., Glover, S. C. O., \& Klessen, R. S.\ 2008, \apj, 672, 757
\bibitem[Clark et al.(2011)]{Clark11} 	Clark, . C., Glover, S. C. O., Klessen, R. S., \& Bromm, V.\ 2011, \apj, 727, 110
\bibitem[Duquennoy \& Mayor(1991)]{Duquennoy91} Duquennoy, A., \& Mayor, M.\ 1991, \aap, 248, 485 
\bibitem[Frebel et al.(2005)]{Frebel05}	Frebel, A., et al. \ 2005,Nature, 434, 871
\bibitem[Greif et al.(2011)]{Greif11} 	Greif, T. H., Springel, V., White, S. D. M., Glover, S. C. O., Clark, P. C., Smith, R. J., Klessen, R. S., \& Bromm, V. \ 2011, \apj, 737, 75
\bibitem[Iwamoto et al.(2005)]{Iwamoto05} Iwamoto, N., Umeda, H., Tominaga, N., Nomoto, K., \& Maeda, K. 2005, Science, 309, 451
\bibitem[Komiya (2011)]{Komiya11}	Komiya, Y. 2011, \apj, 736, 73
\bibitem[Komiya et al.(2009b)]{Komiya09L}	Komiya, Y., Habe, A., Suda, T., \& Fujimoto, Y. M. 2009, \apjl, 696L, 79
\bibitem[Komiya et al.(2010)]{Komiya10}	Komiya, Y., Habe, A., Suda, T., \& Fujimoto, Y. M. 2010, \apj, 717, 542
\bibitem[Komiya et al.(2009a)]{Komiya09}	Komiya, Y., Suda, T., \& Fujimoto, Y. M. 2009, \apj, 694, 1577
\bibitem[Komiya et al.(2007)]{Komiya07}	Komiya, Y., Suda, T., Minaguchi, H., Shigeyama, T., Aoki, W., \& Fujimoto, Y. M. 2007, \apj, 658, 367
\bibitem[Komiya et al.(2013)]{Komiya13}	Komiya, Y., Yamada, S., Suda, T., \& Fujimoto, Y. M. 2013, submitted to \apj
\bibitem[Machida et al.(2008)]{Machida08}	Machida, M. N., Omukai, K., Matsumoto, T., \& Inutsuka, S. \ 2008, \apj, 677, 813
\bibitem[Maynet et al.(2006)]{Maynet06} Meynet, G., Ekstr\"{o}m, S., \& Maeder, A. 2006, \aap, 447, 623
\bibitem[O'Shea et al.(2005)]{OShea05} O'Shea, B. W., Abel, T., Whalen, D., Norman, M. L. \ 2005, \apjl, 628, L5
\bibitem[Rastegaev(2010)]{Rastegaev10} Rastegaev, D. \ 2010, \aj, 140, 2013
\bibitem[Somerville \& Kolatt(1999)]{SK99} Somerville, R.~S., \& Kolatt, T.~S.\ 1999, \mnras, 305, 1 
\bibitem[Stacy et al.(2010)]{Stacy10}	Stacy, A., Greif, T. H., Bromm, V. \ 2010, \mnras, 403, 45
\bibitem[Turk et al.(2009)]{Turk09}	Turk, M. J., Abel, T., O'Shea, B. \ 2009, Science, 325, 601
\bibitem[Uehara \& Inutsuka(2000)]{Uehara00} Uehara, H., \& Inutsuka, S. 2000, \apjl, 531, L91
\bibitem[Umeda \& Nomoto(2003)]{Umeda03} Umeda, H., \& Nomoto, K. 2003, Nature, 422, 871
\bibitem[Yoshida et al.(2003)]{Yoshida03} Yoshida, N., Abel, T., Hernquist, L., Sugiyama, N. 2003, \apj, 592, 645
\bibitem[Wagoner, Fowler, \& Hoyle(1967)]{wagoner} Wagoner, R.~V., Fowler, W.~A., \& Hoyle, F.\ 1967, \apj, 148, 3 
\bibitem[Woosley(2002)]{Woosley02} Woosley, S. E., Heger, A., \& Weaver, T. A. 2002, RvMP, 74, 1015


\end{thebibliography}

\end{document}